\documentclass{sig-alternate-2013}
\usepackage{url}
\usepackage{graphicx}
\usepackage{subfigure}
\usepackage{multirow}
\usepackage{algorithmic}
\usepackage[ruled,vlined]{algorithm2e}

\newfont{\mycrnotice}{ptmr8t at 7pt}
\newfont{\myconfname}{ptmri8t at 7pt}
\let\crnotice\mycrnotice%
\let\confname\myconfname%

\permission{Permission to make digital or hard copies of all or part of this work for personal or classroom use is granted without fee provided that copies are not made or distributed for profit or commercial advantage and that copies bear this notice and the full citation on the first page. Copyrights for components of this work owned by others than ACM must be honored. Abstracting with credit is permitted. To copy otherwise, or republish, to post on servers or to redistribute to lists, requires prior specific permission and/or a fee. Request permissions from permissions@acm.org.}
\conferenceinfo{WSDM'15,}{February 2--6, 2015, Shanghai, China.}
\copyrightetc{Copyright 2015 ACM \the\acmcopyr}
\crdata{978-1-4503-3317-7/15/02\ ...\$15.00.\\
http://dx.doi.org/10.1145/2684822.2685295}

\begin{document}
%

\title{Negative Link Prediction in Social Media}
\numberofauthors{1}
\author{
\alignauthor Jiliang Tang${}^{\ddag}$, Shiyu Chang${}^{\sharp}$, Charu Aggarwal${}^{\dag}$ and Huan Liu${}^{\ddag}$\\
       \affaddr{${}^{\ddag}$Computer Science \& Engineering, Arizona State University}\\
       \affaddr{${}^{\sharp}$ Beckman Institute, University of Illinois at Urbana-Champaign}\\
       \affaddr{${}^{\dag}$ IBM T.J. Watson Research Center}\\
       \email{\{Jiliang.Tang,Huan.Liu\}@asu.edu${}^{\ddag}$,\{chang87\}@illinois.edu${}^{\sharp}$,\{charu\}@us.ibm.com${}^{\dag}$}
}
\maketitle
\begin{abstract}
Signed network analysis has attracted increasing attention in recent
years. This is in part because research on signed network analysis
suggests that negative links have added value in the analytical
process. A major impediment in their effective use is that most
social media sites do not enable  users to specify them explicitly.
In other words, a gap exists between the importance of negative
links and their availability in real data sets.  Therefore, it is
natural to explore whether one can predict negative links
automatically from the commonly available social network data.  In
this paper, we investigate the novel problem of negative link
prediction with only positive links and content-centric interactions
in social media. We make a number of important observations about
negative links, and propose  a principled framework NeLP, which can
exploit positive links and content-centric interactions to predict
negative links. Our experimental results on real-world social
networks  demonstrate that the proposed NeLP framework can
accurately predict negative links with positive links and
content-centric interactions.  Our detailed
experiments also illustrate the relative importance of various
factors to the effectiveness of the proposed framework.
\end{abstract}

\category{H3.3}{Information Storage and Retrieval}{Information Search and Retrieval}[Information filtering]
\terms{Algorithms; Design; Experimentation}

\vfill\eject
\section{Introduction}

Social networks have enabled a vast diversity of relations between
users such as friendships in
Facebook\footnote{\url{https://www.facebook.com/}}, follower
relations in Twitter\footnote{\url{https://twitter.com/}} and trust
relations in Epinions\footnote{\url{http://www.epinions.com/}}. The
increasing availability of large-scale online social network data is
useful not only for  various tasks in social network analysis such
as community detection~\cite{papadopoulos2012community} and link
prediction~\cite{Libe-Klei03}, but it is also leveraged for various
traditional data mining tasks such as feature
selection~\cite{tang2012feature} and
recommendations~\cite{tang2013social}. The vast majority of existing
research has overwhelmingly focused on social networks with only
positive links.  However, social networks can contain  both positive
and negative links. Examples of signed social networks  include
Epinions with trust and distrust links, and
Slashdot\footnote{\url{http://slashdot.org/}} with friend and foe
links. The recent availability of signed social networks in social
media sites such as Epinions and Slashdot  has motivated increasing
research on signed network
analysis~\cite{kunegis2009slashdot,Lesk-etal10,chiang2013prediction}.

It is evident from recent work that negative links have significant
added value over positive links in various analytical tasks. For
example, a small number of negative links can significantly improve
positive link prediction~\cite{Guha-etal04,Lesk-etal10}, and they
can also improve the performance of recommender systems in social
media~\cite{victor2009trust,ma2009learning}.
  Similarly,  trust and distrust relations in Epinions can
help users find high-quality and reliable
reviews~\cite{Guha-etal04}.  Furthermore, the specification of
negative links is interesting in its own right.  On the other hand,
it is generally not very desirable for online social networks to
explicitly collect negative
links~\cite{hardin2004distrust,kunegis2013added}. As a consequence,
most social media sites such as Facebook and Twitter do not enable
users to explicitly specify negative links. Therefore, it is natural
to question whether one can predict negative links automatically
from the available data in social networks. A key assumption is that
while {\em explicit} data is often not available about negative
links, the combination of content-centric and structural data in
social networks may contain implicit information about negative
linkages. While this problem
is very challenging~\cite{chiang2013prediction}, the results of such
an approach  have the potential to improve the quality of the
results of a vast array of applications.

\begin{figure*}[!Ht]
    \begin{center}
      \subfigure[Positive and Negative Link Prediction]{\label{fig:pnlinkprediction}\includegraphics[scale=0.3]{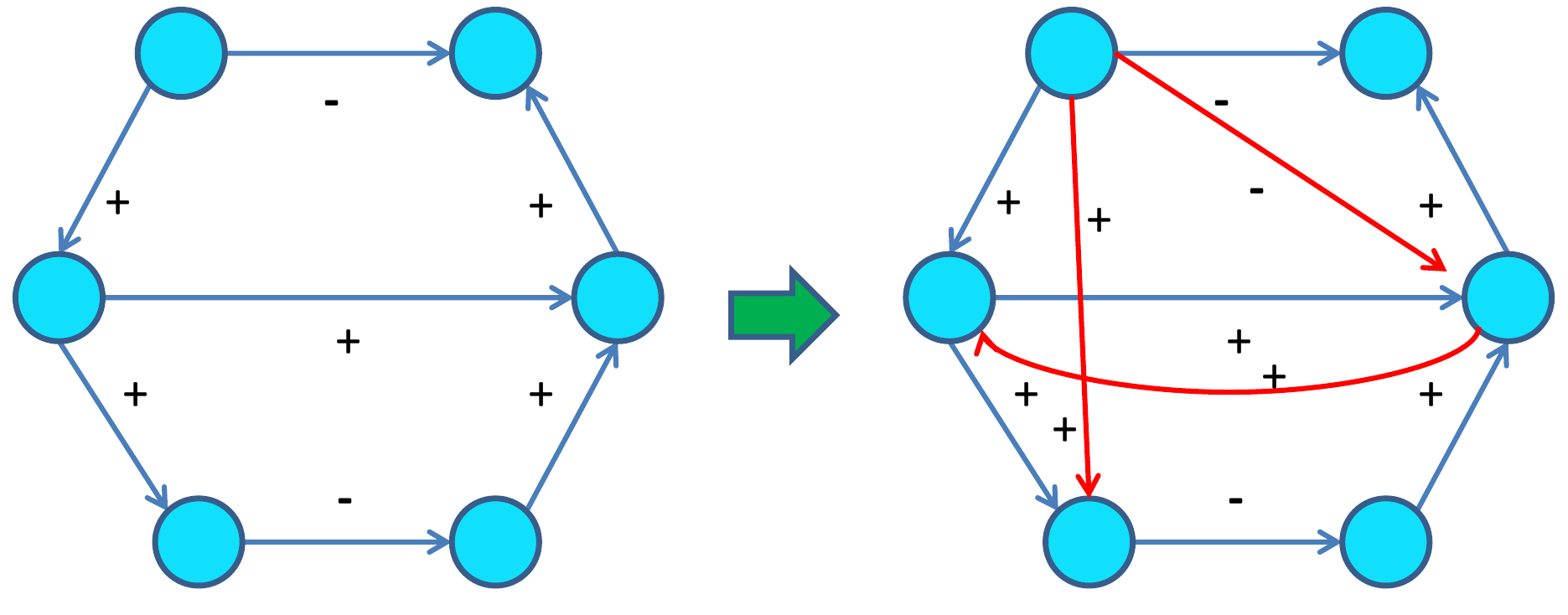}}
    \subfigure[Sign Prediction]{\label{fig:signprediction}\includegraphics[scale=0.3]{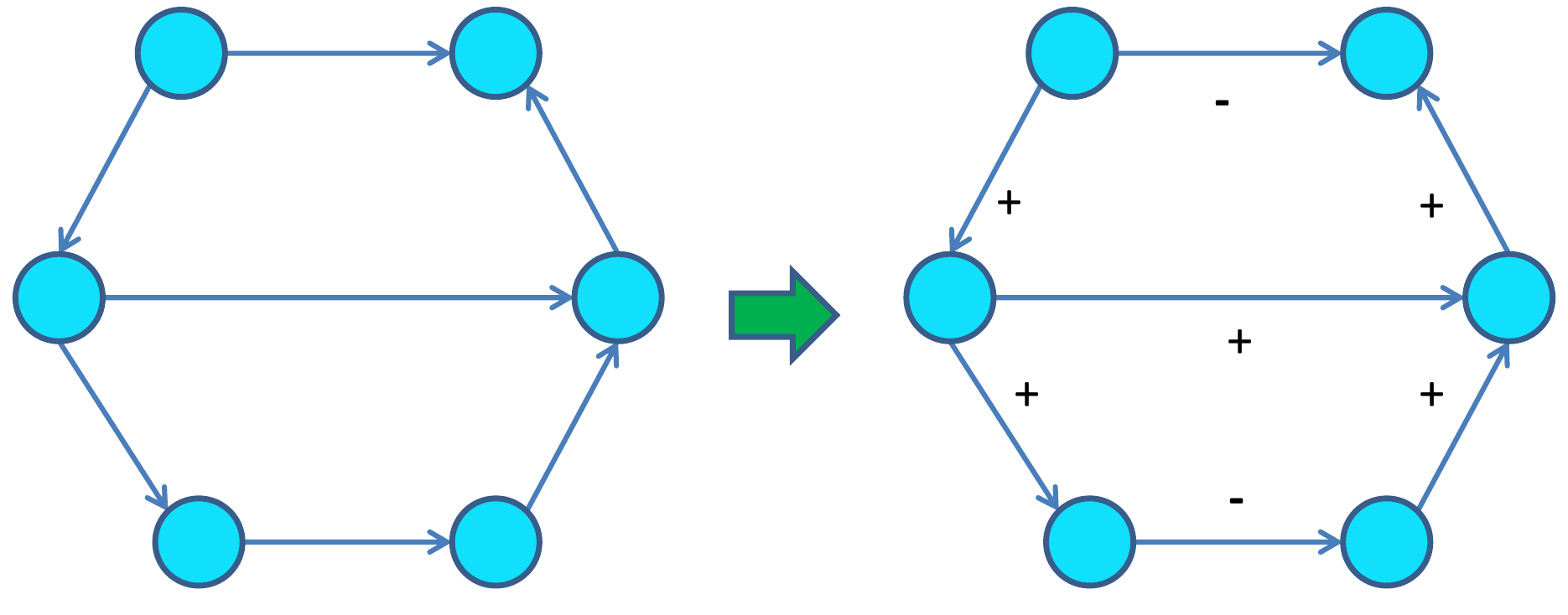}}
    \subfigure[Negative Link Prediction]{\label{fig:nlinkprediction}\includegraphics[scale=0.3]{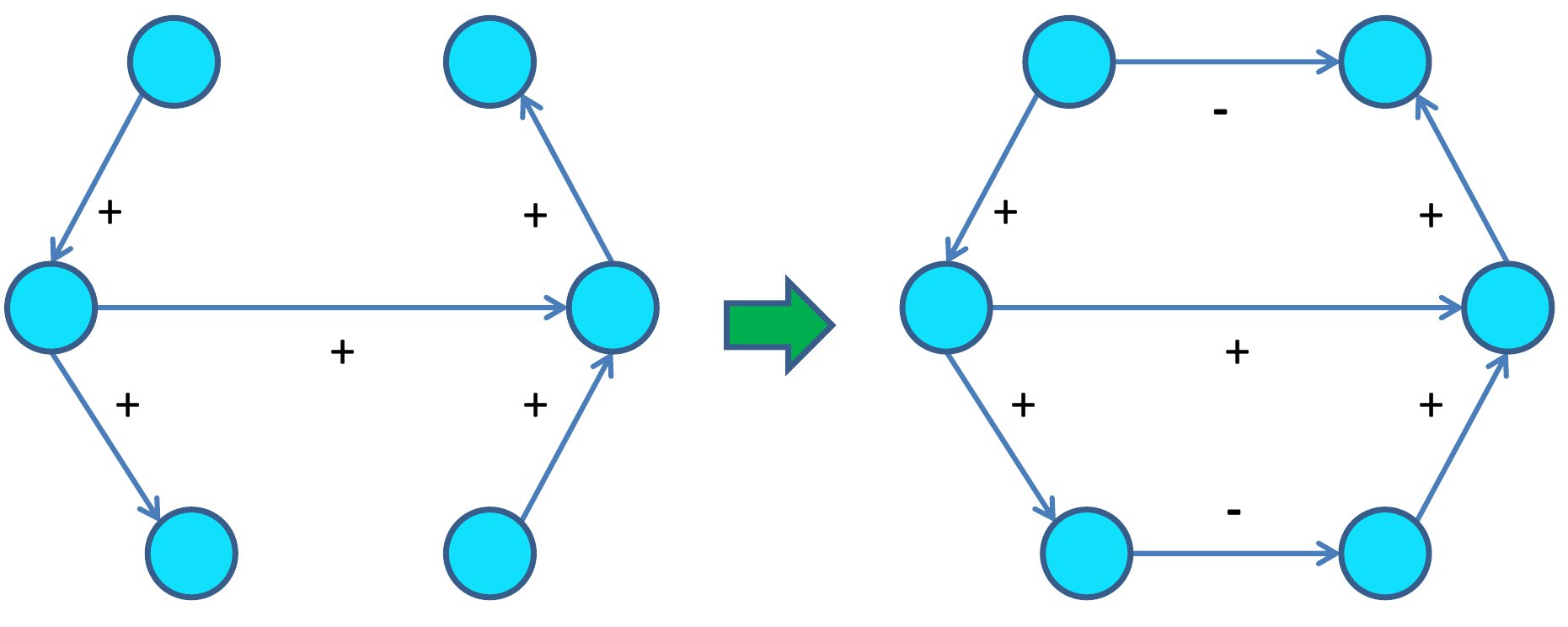}}
    \end{center}
    \vspace*{-0.15in}
\caption{An Illustration of the Differences of Positive and Negative Link Prediction, Sign Prediction and Negative Link Prediction.}
\vspace*{-0.2in}
\label{fig:problemstatement}
\end{figure*}

To preserve the generality of our approach, it is important to use
social network data which is pervasively available across  social
networks in various forms. First, an obvious source of useful data
are the positive linkages which are commonly available in most
social networks.  Second, in most social media sites, users can
create or post content\footnote{ The term ``content'' may be
manifested in diverse ways  such as statuses, tweets, images, or
videos.} and other users can comment, like/dislike and rate such
content. In fact, such user interactions form the dominant  social
media
activities\footnote{\url{http://www.marketingprofs.com/charts/2010/4101/social-media-brand-followers-hunting-for-deals}},
today. For example, users can comment, like, or dislike  content
such posts and videos. Another example is the case of  Epinions,
where users  can rate the helpfulness of reviews written by others.
In this paper, we study the novel problem of negative link
prediction from these  two pervasive sources in social media.
However, we do not assume that examples of  negative links are
available.   To achieve this goal,  we need  (a) theories
explaining the relevance of positive links and content-centric user
interactions for negative link prediction, and (b) combining these
theories with mathematical models to predict negative links. These
goals are achieved by our novel framework for the negative link
prediction problem (NeLP). Our main contributions are summarized
below,
\begin{itemize}
\item We provide  a principled way to exploit positive links and content-centric user interactions for the problem of negative link prediction;
\item We propose a novel framework NeLP to predict negative links by incorporating positive links and content-centric user interactions mathematically; and
\item We evaluate  the proposed NeLP framework in real-world social media datasets to understand the effectiveness and mechanisms
of the proposed framework.
\end{itemize}

The rest of paper is organized as follows. In Section 2, we formally
define the negative link prediction problem. We  perform preliminary
analysis on these datasets in Section 3 to study the interaction
between  existing social theories and negative  link presence. In
Section 4, we combine these theories with a mathematical formulation
for negative link prediction. This is referred to as the NeLP
framework. Section 5 presents experimental results with discussions.
Section 6 briefly reviews related work. Finally, Section 7 concludes
with future work.

\section{Problem Statement}

Let $\mathcal{U} = \{u_1,u_2,\ldots,u_m\}$ be the set of $m$ users
in the social network. A signed network can be decomposed into a
positive network component
$\mathcal{G}_p(\mathcal{U},E_p)$ and a negative network
component $\mathcal{G}_n(\mathcal{U},E_n)$ where
$E_p$ and $E_n$ are the sets of positive and
negative links, respectively. Let $\mathcal{P} =
\{p_1,p_2,\ldots,p_M\}$ be the set of $M$ pieces of content such as posts. We use ${\bf A} \in \mathbb{R}^{m \times
M}$ to denote the user-content relationships  where ${\bf A}_{ij} =
1$ if $p_j$ is created by $u_i$, and ${\bf A}_{ij} = 0$ otherwise.
Users can express opinions on content via comments, likes/dislikes,
and ratings. Some social media sites provide explicit ways of
enabling user feedback on content.  Examples include likes/dislikes
in Youtube, and ``very helpful''/``not helpful'' ratings in
Epinions.   Other more common forms of feedback in large-scale
social networks such as Facebook and Twitter allow users to express
their opinions in the form of textual comments and replies. In such
cases,  we adapt off-the-shelf opinion mining tools to extract user
opinions from such texts. We use ${\bf O} \in \mathbb{R}^{m \times
M}$ to represent the user-post opinion relations where ${\bf O}_{ij}
= 1$, ${\bf O}_{ij} = -1$ and ${\bf O}_{ij} = 0$, if $u_i$ expresses
positive, negative and neutral (or no) opinions, respectively,  on
$p_j$.

With the aforementioned  notations and definitions, the problem of
negative link prediction in social media is formally defined as
follows:\\\\
 {\it Given the positive network $\mathcal{G}_p$, and
content-centric user interactions ${\bf A}$ and ${\bf O}$, we aim to
develop a predictor $f$ to predict the negative network
$\mathcal{G}_n$ with $\mathcal{G}_p$, ${\bf A}$ and ${\bf O}$ as, }
\begin{align}
f:~\{ \mathcal{G}_p, {\bf A}, {\bf O}\} \rightarrow \mathcal{G}_n
\end{align}

The negative link prediction problem in this paper is quite
different and much more challenging than the existing frameworks for
 positive/negative link prediction~\cite{Lesk-etal10},
and the sign prediction problem~\cite{yang2012friend}. An
illustration of their differences  for the existing variations of
the problem is demonstrated in Figure~\ref{fig:problemstatement} as
follows:
\begin{itemize}
\item  One of the existing variations predicts positive and negative links from
existing positive and negative links.  On the other hand, as
illustrated in Figure~\ref{fig:nlinkprediction}, we do not assume
the existence of negative links.
\item The second variation (Figure~\ref{fig:signprediction}) predicts signs of {\em already existing} links.
On the other hand,  the negative link prediction problem needs to
identify the pairs of nodes between which negative links {\em are
predicted to} exist.
\end{itemize}

\section{Data Analysis}

Because positive link prediction is dependent on ``typical''
behavior of social networks such as triadic closure, it is natural
to explore similar properties of negative links with respect to
other positive links, and content-centric interactions. Such an
understanding lays the groundwork for a meaningful negative
link-prediction model. For
the purpose of this study, we collected two datasets from Epinions and Slashdot, that explicitly allow users to express  both
positive and negative links. Note that the  negative links in these
two datasets only serve as a ground-truth about typical properties
and the underlying social theories. However, they are not explicitly
used in the proposed framework for the problem of negative link
prediction.

Epinions is a popular product review site. Users  can create both
positive (trust) and negative (distrust)  links to other users. They
can  write reviews for various products and other users can express
opinions on these reviews with the use of ``helpfulness'' ratings
from 1 to 6. In this work, we view ratings larger than 3 as
positive, and those lower than 3 as negative.  This assumption is
used to populate  the user-content opinion matrix ${\bf O}$  of  the
Epinions dataset.

Slashdot is a technology news platform  in which users can create
friend (positive) and foe (negative) links to other users. They can
also post news articles.  Other users may annotate these articles
with their  comments and opinions. In this case, we computed  the
sentiment polarities of comments based on an off-the-shelf manually
labeled sentiment lexicon, i.e., MPQA Subjectivity
Lexicon~\footnote{\url{http://mpqa.cs.pitt.edu/lexicons/subj$\_$lexicon}}.
These sentiment polarities  are used to populate the user-content
opinion matrix.

Some additional preprocessing was performed on these two datasets by
filtering users without any positive and negative links. A number of
key statistics of these datasets are illustrated  in
Table~\ref{tab:statistics}\footnote{Datasets and code will be available at \url{http://www.public.asu.edu/~jtang20/SignedNetwork.htm}}. It is evident from these statistics that
users are more likely to express positive opinions than negative
opinions in social media.

\begin{table}
\centering
    \vspace*{-0.15in}
\caption{Statistics of the Epinions and Slashdot Datasets.}
\label{tab:statistics}
\begin{tabular}{lcc}
\hline
                                       &Epinions     &Slashdot \\ \hline
\# of Users                            &14,765       &7,275  \\
\# of Positive Links                   &272,513      &67,705\\
\# of Negative Links                   &52,704       &20,851\\
\# of Posts                            &612,321      &300,932\\
\# of Positive Opinions                &6,937,986    &1,742,763\\
\# of Negative Opinions                &163,502      &42,260\\
\hline
\end{tabular}
    \vspace*{-0.15in}
\end{table}

%

\subsection{Where Are our ``Enemies''?}
Our first analytical task is to examine the typical structural
relationships of ``enemies'' within the positive network. In other
words, if $u_i$ has a negative link to $u_j$ in the negative
network $\mathcal{G}_n$, we investigate the typical position of
$u_j$ with respect to $u_i$ in the positive network $\mathcal{G}_p$.
In the following sections, we will use $u_i$+$u_j$, $u_i$-$u_j$
and $u_i$?$u_j$ to denote  positive,  negative and  missing links
between $u_i$ to $u_j$, respectively.

\begin{figure}
    \begin{center}
      \subfigure[Epinions ]{\label{fig:epinions_path}\includegraphics[scale=0.28]{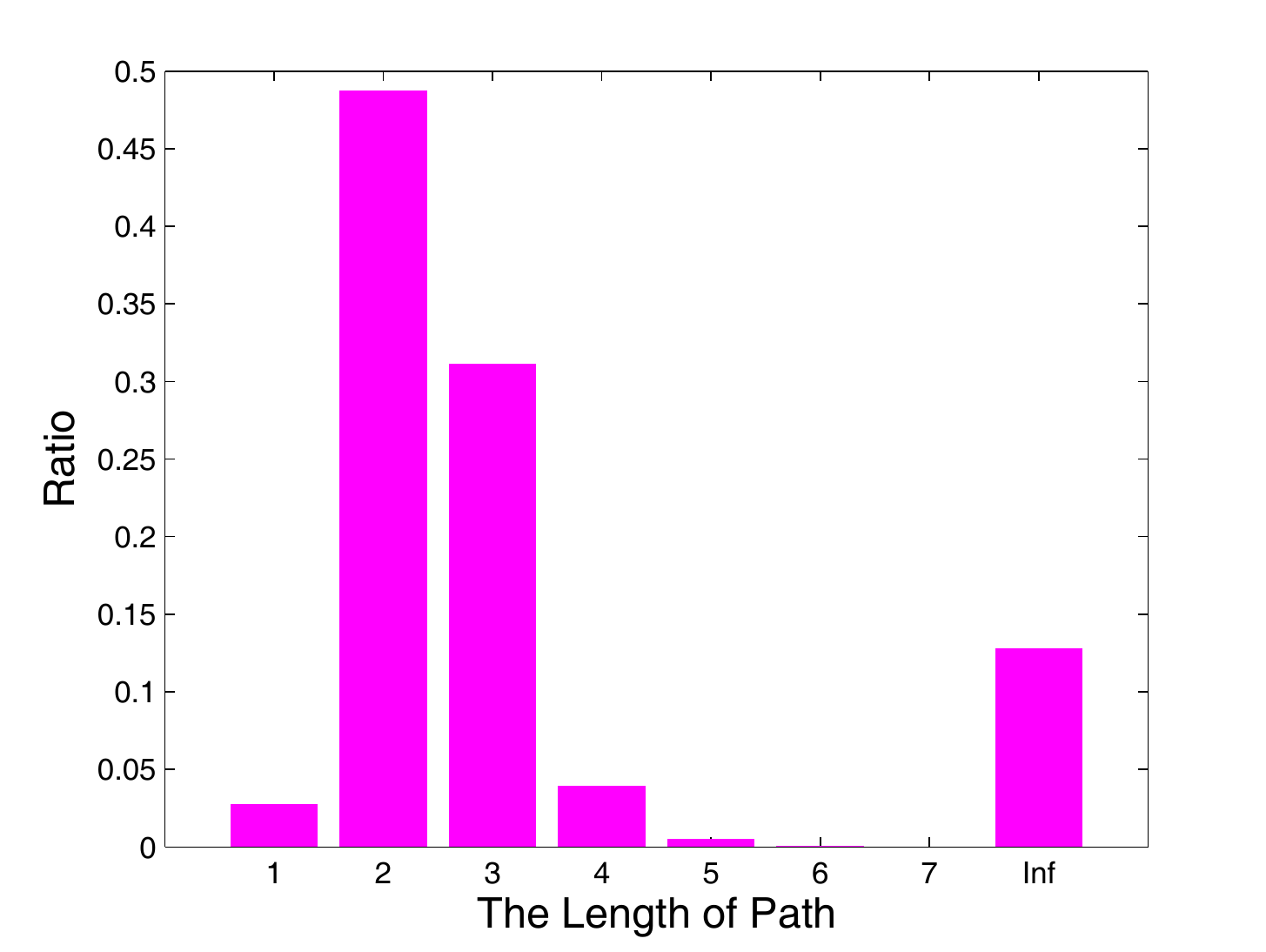}}
    \subfigure[Slashdot]{\label{fig:slash_path}\includegraphics[scale=0.28]{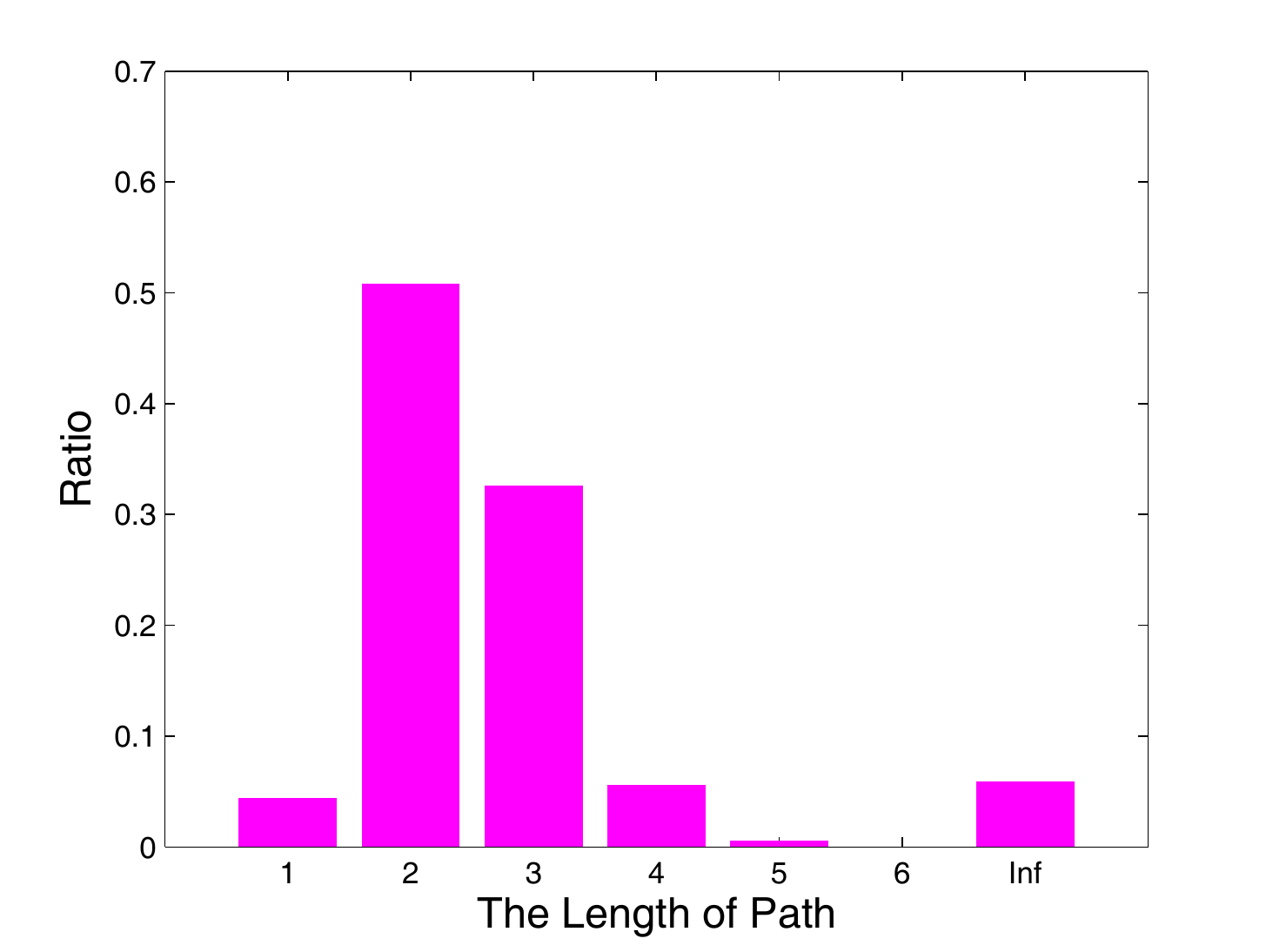}}
    \end{center}
     \vspace*{-0.15in}
\caption{Ratio Distributions of the Length of Shortest Path for Pairs with Negative Links in the Positive Networks.}
    \vspace*{-0.15in}
\label{fig:path}
\end{figure}

For each negative link $u_i$-$u_j$ in $\mathcal{G}_n$, we use
breadth-first search  to compute the shortest path from $u_i$ to
$u_j$ in $\mathcal{G}_p$. If paths exist  from $u_i$ to $u_j$, we
report the length of the shortest path. Otherwise we report the
length as ``inf'' to indicate there is no path from $u_i$ to $u_j$
in $\mathcal{G}_p$. The ratio distributions of the lengths of the
shortest paths for all negative links are demonstrated in
Figures~\ref{fig:epinions_path} and \ref{fig:slash_path} for
Epinions and Slashdot, respectively. In both datasets, more than
$45\%$ of negative links $u_i$-$u_j$ have shortest path lengths less
than $3$, and more than $80\%$ of them have shortest path lengths
less than $4$. These results suggest that our ``enemies'' are often
close to us in the positive network $\mathcal{G}_p$. For example,
about $82.64\%$ and $87.86\%$ of enemy-pairs are  within $3$-hops of
each other in the positive networks of Epinions and Slashdot,
respectively.

\subsection{Social Theories in Signed Networks}

In this subsection, we investigate two of the most important social
theories for signed networks, i.e., balance
theory~\cite{heider1946attitudes,Cart-etal56} and status
theory~\cite{Lesk-etalchi10}.

In general, balance theory is based on the intuition that ''the
friend of my friend is my friend'' and ''the enemy of my enemy is my
friend''~\cite{heider1946attitudes}.  This theory relates  the
balance of signs on a triad involving three users in a social
network with positive and negative links. We use $s_{ij}$ to denote
the sign of the link between $u_i$ and $u_j$ where $s_{ij} =  1$ (or
$s_{ij} = -1$) if there is a positive link (or a negative link)
between $u_i$ and $u_j$. Balance theory suggests that a triad
$\langle u_i, u_j, u_k \rangle$ is balanced if the following is
true - (a) $s_{ij} = 1$ and $s_{jk}=1$, then $s_{ik} = 1$ ; or (b) $s_{ij}=-1$ and $s_{jk}=-1$, then $s_{ik} = 1$.

For a triad $\langle u_i, u_j, u_k \rangle$, there are four possible
signed combinations (+,+,+), (+,+,-), (+,-,-) and (-,-,-), while
only (+,+,+) and (+,-,-) are balanced. Based on the discussion
in~\cite{Lesk-etal10}, the directions of links are ignored in the
study of  balance theory because  balance theory is designed  for
undirected networks. We computed the relative presence of these four
possible combinations and find that $92.31\%$ and $93.01\%$ of
triads in Epinions and Slashdot are balanced, respectively.


In status theory, a positive link from $u_i$ to $u_j$ indicates that
$u_i$ has a higher status than $u_j$; while a negative link from
$u_i$ to $u_j$ indicates that $u_i$ has a lower status than $u_j$.
For a triad, status theory suggests that if we take each negative
link, reverse its direction, and flip its sign to positive, then the
resulting triangle (with all positive link) should be acyclic. We
first obtain all triads and then follow the above way to examine
whether these triads satisfy status theory or not. We find that
$94.73\%$ and $93.38\%$ of triads in Epnions and Slashdot satisfy
status theory, respectively.

\subsection{Negative Links and Content-centric Interactions}

Content-centric  interactions relate the opinion of user $u_i$ on
the content posted by user $u_j$. The user $u_i$ can express
negative opinions on content posted by another user  $u_j$ by
disliking, giving negative comments, or negative ratings. Such types
of  content-centric  interactions may be viewed as negative
interactions between $u_i$ and $u_j$. A negative interaction from
$u_i$ to $u_j$  is often a manifestation of user $u_i$'s
disagreement and antagonism toward $u_j$.  It is therefore
reasonable to surmise that negative interactions might be correlated
with  negative links. In this subsection, we study the correlation
between negative interactions and negative links.

Let ${\bf N}\in\mathbb{R}^{m \times m}$ be a user-user negative
interaction matrix where ${\bf N}_{ij}$ denotes the number of
negative interactions from $u_i$ to $u_j$. We can obtain ${\bf N}$
from the user-content authorship matrix ${\bf A}$ and the
user-content opinion matrix ${\bf O}$ as ${\bf N} = -{\bf A}({\bf
O}^-)^\top$ where ${\bf O}^- = \frac{{\bf O} - |{\bf O}|}{2}$ is the
negative part of ${\bf O}$. To verify the correlation between
negative interactions and negative links, we aim to answer the
following question: Are pairs of users with negative interactions
more likely to have negative links than those without negative
interactions?

For each pair $\langle u_i,u_j \rangle$ with negative interactions
(or ${\bf N}_{ij} \neq 0$), we first randomly select a user $u_k$
that $u_i$ does not have negative interactions with (or ${\bf
N}_{ij} = 0$), and then use $S$ (or $R$) to indicate whether $\langle
u_i,u_j \rangle$ (or $\langle u_i,u_k \rangle$) has a negative link where $S=1$ (or $R=1$) if $u_i$ has a negative link to
$u_j$ (or $u_i$ has a negative link to $u_k$), otherwise $S=0$ (or
$R=0$). Let ${\bf s}$ be a vector of $S$s over all pairs of users
with negative interactions and ${\bf r}$ be the corresponding vector
of $R$s. We conduct a two-sample $t$-test on ${\bf s}$ and
${\bf r}$. The null hypothesis and the alternative hypothesis are
defined as $H_0: {\bf s} \leq {\bf r},~~~H_1: {\bf s} > {\bf r}$.
The null hypothesis is rejected at significance level $\alpha =
0.01$ with p-values of 5.72e-89 and 1.93e-109 for Epinions and
Slashdot, respectively. Evidence from the $t$-test suggests a
positive answer to the question: {\it there is a strong correlation
between negative interactions and negative links, and users with
negative interactions are likely to have negative links.}

\begin{figure}
    \begin{center}
      \subfigure[Epinions ]{\label{fig:epinions_inter}\includegraphics[scale=0.28]{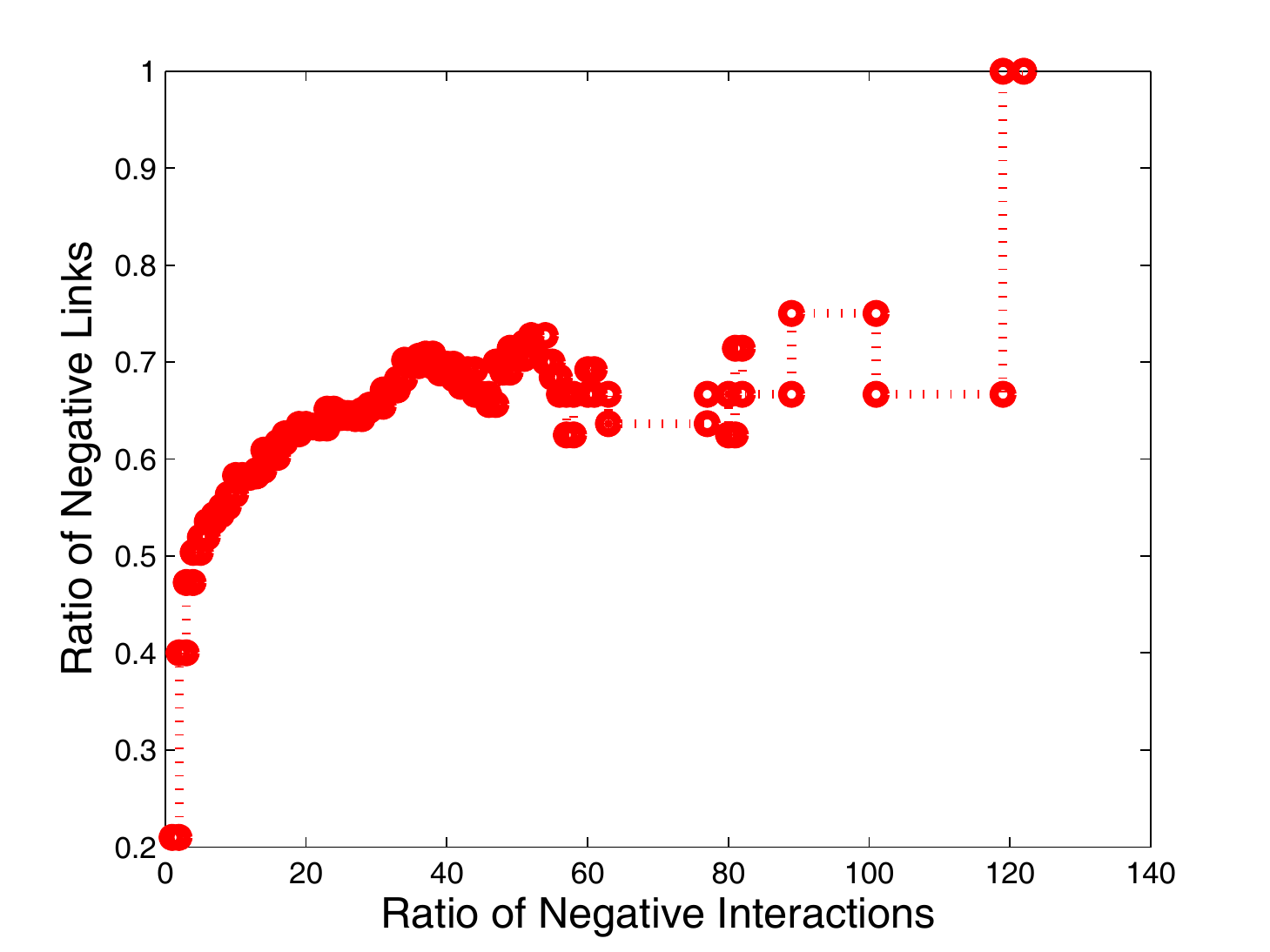}}
    \subfigure[Slashdot]{\label{fig:slash_inter}\includegraphics[scale=0.28]{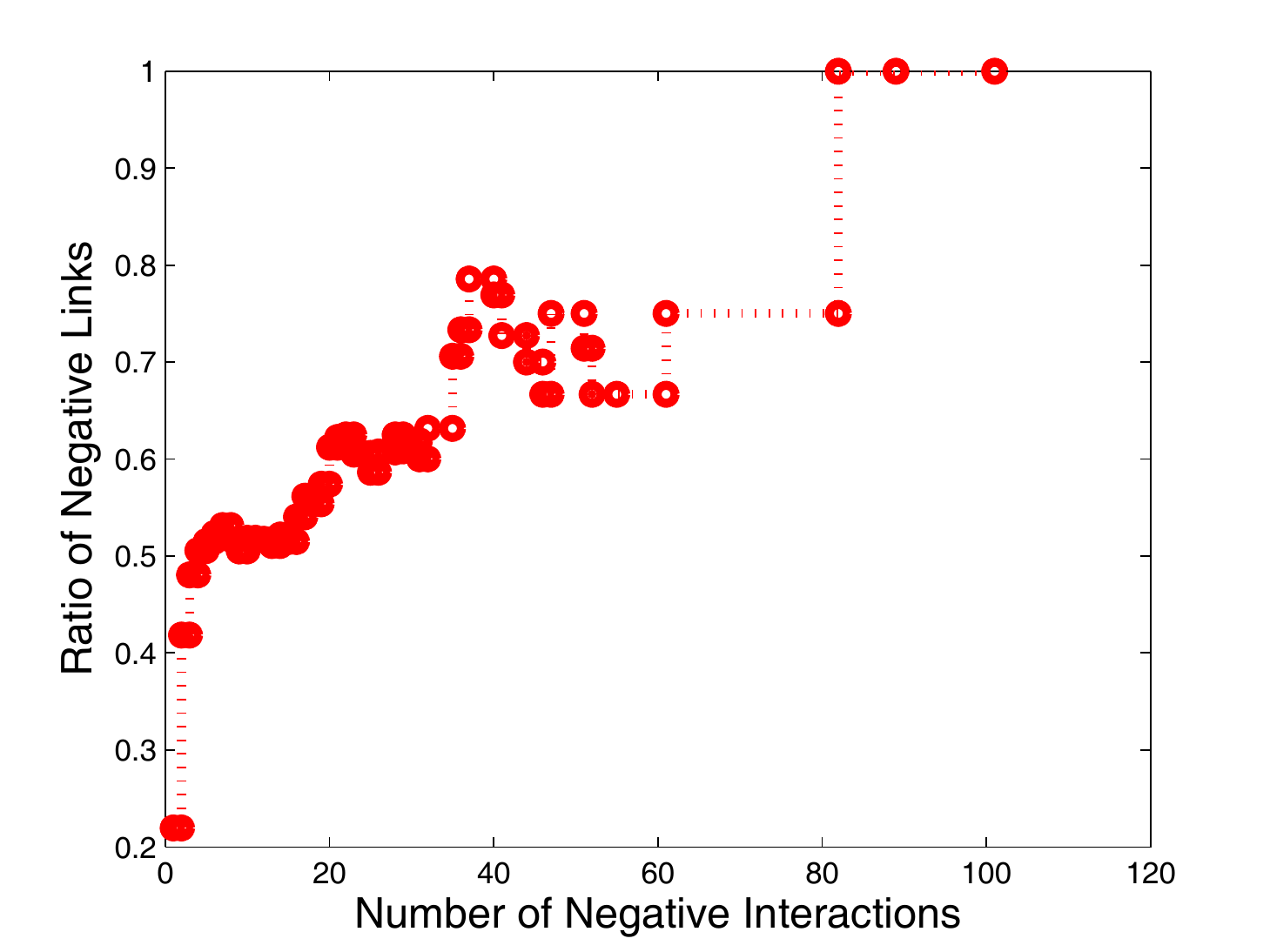}}
    \end{center}
       \vspace*{-0.15in}
\caption{The Ratios of Negative Links with respect to the Number of Negative Interactions.}
    \vspace*{-0.2in}
\label{fig:nncorrelation}
\end{figure}

We further investigate the direct impact of negative interactions on
negative links. For a given value of $K$, we calculated the ratio of
pairs with both negative links and at least $K$ negative
interactions over all pairs with at least $K$ negative interactions.
The ratio distributions with respect to the number of negative
interactions are demonstrated in Figures~\ref{fig:epinions_inter}
and \ref{fig:slash_inter}, respectively. Note that the ratios of
randomly selected pairs with negative links among all $n(n-1)$ pairs
of users are 2.4177e-04 and 3.9402e-04 in Epinions and Slashdot,
respectively. From the figures, we note that the ratios are much
higher than the random ones even when $K$ is very small.  This
observation  further supports the existence of the correlation
between negative interactions and negative links. Furthermore with
increase of $K$, the ratios tend to increase. Therefore, an increase
in the number of negative interactions increases the likelihood of
negative links between users.

\subsection{Discussion}
We summarize the insights obtained in the aforementioned discussion
as follows:
\begin{itemize}
\item Most of our ``enemies'' are close to us within a few (e.g., 2 or 3) hops in the positive
network.
\item Most of triads in signed networks satisfy balance theory and status
theory.
\item Pairs of users with negative interactions are more likely to have negative links than those
without them.
\item  Negative interactions between users increase the propensity of
negative links.
\end{itemize}

These observations provide the groundwork for our proposed framework
for negative link prediction.

\section{The Proposed Framework - N\lowercase{e}LP}

Algorithms for all variations of the link prediction problem are
either unsupervised methods~\cite{Libe-Klei03,hsieh2012low} or
supervised methods~\cite{lichtenwalter2010new,Lesk-etal10}.
Supervised methods consider the link prediction problem as a
classification problem by using the existence of links as labels and
have several advantages over unsupervised methods such as superior
performance, adaptability  to different data domains, and variance
reduction~\cite{lichtenwalter2010new}. Similar to traditional
supervised link prediction, we can consider the negative link
prediction problem as a classification problem where we need to
construct labels and extract features. Different from traditional
link prediction, there are unique challenges in preparing
training data in the negative link prediction problem. For example, existing links are given in traditional link
prediction such as positive links in positive link prediction, and positive and negative links in positive and negative link prediction, while existing negative links are not given in the negative link prediction problem. 
Next we will first give details about label
construction and feature extraction for the negative link prediction
problem. Finally, the optimization formulation and solution will be
discussed.

\subsection{Label Construction}

Let $E_o$ denote pairs of users with missing links. In most previous formulations
of link prediction, including the signed version, label
construction is trivial because the presence of links is specified. However, we study the scenario where no
negative links are provided, and therefore the labels for
$E_n$ are really an unspecified subset of $E_o \cup E_n$.
What is worse, the sizes of $E_n$ and $E_o$ are
extremely imbalanced. For example, the imbalance ratios
$E_n:E_o$ are 1:4131 and 1:2534 in Epinions and
Slashdot respectively. We treat missing links as positive samples and negative links as negative samples. Label construction is to construct positive and negative samples from $E_o \cup E_n$. Since the ratio of $E_o$ in
$E_n \cup E_o$ are often bigger than $99.9\%$,
we can randomly select a subset of samples from $E_n \cup
E_o$ as positive samples $PS$. Next we introduce a way to
select samples from $E_n \cup E_o$ as negative samples based on previous observations.
The process is shown in Algorithm~\ref{alg:negativesample}.

\begin{algorithm}
{\bf Input :} The positive network $\mathcal{G}_p$ and user-user negative interaction matrix ${\bf N}$ \\
{\bf Output :} Negative sample set $NS$ and the reliability weight matrix ${\bf W}$

\begin{algorithmic}[1]
\STATE Initialize $NS = \emptyset$ \FORALL{${\bf N}_{ij} \neq 0$}
\STATE $NS = NS \cup \{\langle u_i,u_j \rangle\}$ \ENDFOR \STATE
Construct $\mathcal{G}$ as a signed network with positive links from
${\bf G}_p$ and negative links $u_i$-$u_j$ from $NS$ \STATE Remove
samples $\langle u_i,u_j \rangle$ from $NS$ if $u_i$-$u_j$ is in any
triads of $\mathcal{G}$ that does not satisfy status theory \STATE
Add samples $\langle u_i,u_k \rangle$ into $NS$ if $u_i$-$u_k$ can
make all triads that involve $u_i$ and $u_k$ in $\mathcal{G}$
satisfying status theory
\FORALL{ $\langle u_i,u_j \rangle \in NS$}
\STATE Calculate a reliability weight ${\bf W}_{ij}$
\ENDFOR
\end{algorithmic}
\caption{Negative Sample Construction}
\label{alg:negativesample}
\end{algorithm}

Next, we describe  Algorithm~\ref{alg:negativesample} for negative
sample construction. The strong correlation between negative
interactions and negative links suggests that users with negative
interactions are likely to have negative links. Therefore from line
2 to line 4 in Algorithm~\ref{alg:negativesample}, we construct
negative sample candidates based on this observation. With the
positive links from $\mathcal{G}_p$ and negative links $u_i$-$u_j$
from $NS$, we construct a signed network $\mathcal{G}$ in line 5.
Most of the triads in signed networks satisfy status theory.
Therefore we refine $NS$ by (a) excluding $\langle u_i,u_j\rangle$
from $NS$ if $u_i$-$u_j$ is in any triads of $\mathcal{G}$ that does
not satisfy status theory in line 6; and (b) adding samples $\langle
u_i,u_k \rangle$ into $NS$ if $u_i$-$u_k$ can make all triads that
involve $u_i$ and $u_k$ in $\mathcal{G}$ satisfying status theory in
line 7. The reliability of these negative samples may vary. For
example, observations from data analysis indicate that negative
sample candidates with more negative interactions are more likely to
have negative links, and are therefore  more likely to be  reliable.
Therefore, we associate each $\langle u_i,u_j \rangle$ with a
reliability weight ${\bf W}_{ij}$, which is defined as follows:
\begin{align}
{\bf W}_{ij} =\left\{
\begin{array}{l}
f({\bf N}_{ij})~~~~~~~~~~~~~~~\text{if ${\bf N}_{ij} \neq 0$} \\
r~~~~~~~~~~~~~~~~~~~~~~\text{otherwise} \\
\end{array}
\right. .
\label{eq:dw}
\end{align}
\noindent if the pair $\langle u_i,u_j \rangle \in NS$ has negative
interactions, we define the reliability weight as a function $f$ of
the number of negative interactions ${\bf N}_{ij}$ where $f(x) \in
[0,1]$ is a non-decreasing function of $x$. This is because  the
more negative interactions two users have, the more likely it is
that a negative link exists between them. Otherwise, the pair $
\langle u_i,u_j\rangle \in NS$ is added  by line 7 in
Algorithm~\ref{alg:negativesample} and we set the reliability weight
to a constant $r$.

\subsection{Feature Extraction}

We extract three types of features corresponding to user features,
pair features and sign features. User features and pair features are
extracted from two given sources, such as positive links and
content-centric interactions, as follows:
\begin{itemize}
\item User features are extracted for each user $u_i$
including $u_i$'s indegree (or outdegree) in terms of positive
links, the number of triads that $u_i$ involved in, the number of
content-centric items (e.g., posts) that $u_i$ creates, the number
of $u_i$'s posts that obtain positive (or negative) opinions, and
the number of positive (or negative) opinions $u_i$ expresses; and
\item Pair features are extracted for a pair of users $\langle u_i,u_j\rangle$ including
 the number of positive (or negative) interactions from $u_i$ to $u_j$, the number of positive (or negative)
 interactions from $u_j$ to $u_i$, Jaccard coefficients of indegree (or outdegree) of $u_i$ and $u_j$ in terms of positive links,
 and the length of the shortest path between $u_i$ and $u_j$.
\end{itemize}

We construct a weighted signed network with the given positive links and negative links from $NS$ where the weights of positive
links are 1 and the weights of negative links are their reliability weights.
 For a pair $\langle u_i,u_j \rangle$, signed features include weighted indegree (or outdegee) in terms of negative links of $u_i$, weighted indegree
 (and outdegee) in terms of negative links of $u_j$, Jaccard coefficients of indegree (or outdegree) of $u_i$ and $u_j$
  in terms of negative links and $16$ weighted triads suggested by~\cite{Lesk-etal10}.

With definitions of user features, pair features and sign features,
we extract $45$ features in total for each pair $\langle
u_i,u_j\rangle$ including $8$ user features of $u_i$, $8$ user
features of $u_j$, $7$ pair features, and $22$ signed features.

\subsection{The N\lowercase{e}LP Optimization Framework}

Through label construction and feature extraction, we prepare
training data to learn classifiers for the negative link prediction
problem. However, the  labels of the training data are noisy, and
especially so for negative samples. Therefore, it is necessary for
the base classifier to be tolerant to training data noise.  In
this paper, we choose a soft-margin version of support vector
machines as our basic classifier because it has been proven to be
highly noise-tolerant~\cite{cristianini2000introduction}.

Let $\mathcal{X} = \{x_1,x_2,\ldots,x_N\}$ be the set of user pairs
in $E_o \cup E_n$ and ${\bf x}_i$ be the feature
vector representation of the pair $x_i$. The standard soft-margin
support vector machine for the negative link prediction problem is
as follows:
\begin{align}
&\min_{{\bf w},b,\epsilon}~~~\frac{1}{2} \|{\bf w}\|_2^2 + C \sum_{x_i \in PS \cup NS } \epsilon_i \nonumber \\
&~~~s.t.~~~ y_i ( {\bf w}^\top {\bf x}_i + b) \geq 1 - \epsilon_i,~ x_i \in PS \cup NS \nonumber \\
&~~~~~~~~~~\epsilon_i \geq 0 ~~ x_i \in PS \cup NS
\label{eq:standard}
\end{align}

Eq.~(\ref{eq:standard}) introduces the term  $\epsilon_i$  for the
soft-margin slack variable of $x_i$, which
can be viewed as the allowance for the noise in this training
sample. The parameter $C$ controls the degree of impact of this
term. In the negative link-prediction problem, the noise-levels of
positive and negative samples are different because positive samples
$PS$ are generally more robust than the (indirectly derived)
negative samples. As discussed earlier, the reliability of negative
samples is explicitly quantified with their weights.  These
intuitions suggest that we should allow more errors in negative
samples especially when their weights suggest unreliability. This
yields the following formulation:
\begin{align}
&\min_{{\bf w},b,\epsilon}~~~\frac{1}{2} \|{\bf w}\|_2^2 + C_p \sum_{x_i \in PS } \epsilon_i + C_n \sum_{x_j \in NS} c_j \epsilon_j \nonumber \\
&~~~~~~s.t.~~~~~~~ y_i ( {\bf w}^\top {\bf x}_i + b) \geq 1 - \epsilon_i, ~~ x_i \in PS \nonumber \\
&~~~~~~~~~~~~~~~~~~ y_j ( {\bf w}^\top {\bf x}_j + b) \geq 1 - \epsilon_j,~~ x_j\in NS\nonumber \\
&~~~~~~~~~~~~~~~~~~ \epsilon_i \geq 0,~\epsilon_j \geq 0
\label{eq:biased}
\end{align}

In Eq.~(\ref{eq:biased}), we use two parameters $C_p$ and $C_n$ to
weight the  positive and negative errors differently.  We use a
larger value for $C_p$ compared to  $C_n$ to reflect the
differential behavior of the positive and negative samples. For a
negative sample $x_j$, we introduce a weight $c_j$ to further
control its error based on its quantified reliability weight. For
the negative sample $x_j$ corresponding to  the pair $\langle
u_i,u_k\rangle$, we set  $c_j = {\bf W}_{ik}$ where ${\bf W}_{ik}$
is the reliability weight for $\langle u_i,u_k\rangle$. This
additional term allows differential control of the noise in negative
samples of varying reliability.

Balance theory suggests that triads in signed networks are likely to
be balanced. Therefore, if there is a positive link between $u_i$
and $u_j$, and both $u_i$ and $u_j$ do not have positive links with
another user $u_k$, the types of $(u_i,u_k)$ and $(u_j,u_k)$ in the
negative graph $\mathcal{G}_n$ are likely to be the same. In other
words, to ensure the structural balance, it is likely that both are negative links where $\langle u_i, u_j, u_k \rangle$ forms a balanced triad or both are missing links where there is no triad among $\langle u_i, u_j, u_k \rangle$. With this intuition, we introduce a matrix ${\bf B}$ where ${\bf
B}_{h\ell}=1$ if there is a positive link between $u_i$ and $u_j$
where we assume that $x_h$ and $x_\ell$ denote pairs $\langle
u_i,u_k\rangle$ and $\langle u_j,u_k\rangle$ respectively.
Otherwise, we assume that ${\bf B}_{h\ell}=0$. Then, we force $x_h$
and $x_\ell$ to have the same types of links if ${\bf B}_{h\ell}=1$
by introducing a balance-theory regularization:
\begin{align}
\min~~\frac{1}{2}\sum_{h,\ell} {\bf B}_{h\ell} ({\bf w}^\top {\bf x}_h - {\bf w}^\top {\bf x}_\ell)_2^2 = {\bf w}^\top {\bf X} \mathcal{L} {\bf X}^\top {\bf w}
\label{eq:btr}
\end{align}
\noindent  Here,  $\mathcal{L}$ is the Laplacian
matrix based on ${\bf B}$. The number of pairs in $E_n \cup E_o$ is usually very large, which leads to a large number
of terms in the balance theory regularization. The observation from
data analysis suggests that our ``enemies'' are usually close to us
in the positive network.  Hence, in this work, we only consider
pairs whose shortest path lengths are $2$, and pairs in $NS$ and
$PS$ in the balance theory regularization. We assume that there are $l+\mu$ samples in ${\bf X}$ where the first $l$ ones are from $PS \cup NS$.
The significance of the introduction of the balance theory regularization is two-fold.
First, it allows us to model balance theory. Second, it allows us to
include more samples during the learning process in addition to $NS$
and $PS$. A similar function is achieved by this approach, as
achieved by unlabeled samples in semi-supervised
learning~\cite{zhu2003semi}. With these components, the proposed
NeLP framework is able to solve the following optimization problem:
\begin{align}
&\min_{{\bf w},b,\epsilon}~~~\frac{1}{2} \|{\bf w}\|_2^2 + C_p \sum_{x_i \in PS } \epsilon_i + C_n \sum_{x_j \in NS} c_j \epsilon_j \nonumber \\
&~~~~~~~~~~~~~+\frac{C_b}{2} {\bf w}^\top {\bf X} \mathcal{L} {\bf X}^\top {\bf w} \nonumber \\
&~~~~~~s.t.~~~~~~~ y_i ( {\bf w}^\top {\bf x}_i + b) \geq 1 - \epsilon_i, ~~ x_i \in PS \nonumber \\
&~~~~~~~~~~~~~~~~~~ y_j ( {\bf w}^\top {\bf x}_j + b) \geq 1 - \epsilon_j,~~ x_j\in NS\nonumber \\
&~~~~~~~~~~~~~~~~~~ \epsilon_i \geq 0,~\epsilon_j \geq 0
\label{eq:original}
\end{align}

We solve the optimization problem in Eq.~(\ref{eq:original}) based
on the dual form~\cite{belkin2005manifold}. The classical
representer theorem states that the solution to this minimization
problem of Eq.(~\ref{eq:original}) exists in $\mathcal{H}_K$ and can
be written as follows:
\begin{align}
{\bf w}^* = \sum_i \alpha_i K({\bf x}_i, {\bf x})
\end{align}

Eq.(~\ref{eq:original}) can be rewritten as follows:
\begin{align}
&\min_{\alpha, b, \epsilon}~~\frac{1}{2}\alpha^\top {\bf K} \alpha + C_p \sum_{u_i \in PS} \epsilon_i + C_n \sum_{u_j\in NS} c_j \epsilon_j + \frac{C_b}{2} \alpha^\top {\bf K} \mathcal{L} {\bf K} \alpha\nonumber \\
&~~~~~~s.t.~~~~~~~ y_i ( \sum_k \alpha_k K({\bf x}_k, {\bf x}_i) + b) \geq 1 - \epsilon_i, ~~ u_i \in PS \nonumber \\
&~~~~~~~~~~~~~~~~~~ y_j ( \sum_k \alpha_k K( {\bf x}_k, {\bf x}_j) + b) \geq 1 - \epsilon_j,~~ u_j\in NS\nonumber \\
&~~~~~~~~~~~~~~~~~~ \epsilon_i \geq 0,~\epsilon_j \geq 0
\label{eq:representer}
\end{align}
\noindent where ${\bf K}$ is the Gram matrix over all samples.

We define $s_i$ for $x_i$ as follows:
\begin{align}
s_i =
\left\{
\begin{array}{l}
C_p \quad \quad \text{for $x_i \in  PS$,}\\
C_n c_i \quad   \text{for $x_i \in  NS$}.\\
\end{array}
\right.
\end{align}

After the introduction of two sets of multipliers $\beta$ and
$\gamma$, the Lagrangian function of Eq.(~\ref{eq:representer}) is
as follows:
\begin{align}
& L({\bf w},b,\epsilon,\alpha,\gamma) = \frac{1}{2}\alpha^\top ({\bf K} + C_b {\bf K} \mathcal{L} {\bf K}) \alpha + \sum_{i=1}^l s_i \epsilon_i \nonumber \\
& - \sum_{i=1}^l \beta_i [y_i ( \sum_k \alpha_k K({\bf x}_k, {\bf x}_i) + b) - 1 + \epsilon_i] - \sum_{i=1}^l \gamma_i \epsilon_i
\end{align}
\noindent where $\beta$ and $\gamma$ are Lagrange multipliers.

To obtain the dual representation, we set
\begin{align}
&\frac{\partial{L}}{\partial{b}} = 0  \Rightarrow \sum_{i=1}^l \beta_i y_i = 0 \nonumber \\
&\frac{\partial{L}}{\partial{\epsilon_i}} = 0 \Rightarrow s_i - \beta_i - \gamma_i = 0 \Rightarrow 0 \leq \beta_i \leq s_i
\label{eq:dual}
\end{align}

With Eq.~(\ref{eq:dual}), we can rewrite the Lagrangian as a
function of only $\alpha$ and $\beta$ as follows:
\begin{align}
L(\alpha, \beta) = \frac{1}{2}\alpha^\top ({\bf K} + C_b {\bf K} \mathcal{L} {\bf K}) \alpha - \alpha^\top {\bf K} {\bf J}^\top {\bf Y} \beta + \sum_{i=1}^l \beta_i
\label{eq:middle}
\end{align}
\noindent in Eq.~(\ref{eq:middle}), ${\bf J} = [{\bf I}~{\bf 0}]$
where ${\bf I}$ is an $l\times l$ identity matrix and ${\bf 0}$ is a $l\times\mu$ rectangular
matrix with all zeros, and ${\bf Y}$ is a $l\times l$ diagonal matrix composed
by labels of samples in $PS$ and $NS$.

By setting $\frac{\partial{L}}{\partial{\alpha}} = 0$, we obtain
\begin{align}
\alpha = ({\bf I} + C_b {\bf K} \mathcal{L} )^{-1} {\bf J}^\top {\bf Y} \beta
\end{align}

After substituting back in the Lagrangian function, we obtain the
dual problem as a quadratic programming problem:
\begin{align}
&\max_{\beta}~~\sum_{i=1}^l \beta_i - \frac{1}{2}\beta^\top {\bf Q} \beta \nonumber \\
&~~~~~~~~s.t.~~~~~~~\sum_{i=1}^l \beta_i y_i = 0 \nonumber \\
&~~~~~~~~~~~~~~~~~~~0 \leq \beta_i \leq s_i
\end{align}
\noindent where ${\bf Q}$ is defined as follows:
\begin{align}
{\bf Q} = {\bf Y} {\bf J} {\bf K}({\bf I} + C_b {\bf K} \mathcal{L} )^{-1} {\bf J}^\top {\bf Y}
\end{align}

%
%
%

\section{Experiments}

In this section, we present experiments which (a) quantify the
performance of the proposed  NeLP framework in predicting negative
links, and (b) evaluate the contribution of various model components
to the performance.  We begin by introducing performance  evaluation
metrics, which are useful in both contexts.

\subsection{Evaluation Metrics}

All forms of link prediction can be viewed as highly imbalanced
classification problems. In such cases, straightforward accuracy
measures are well known to be misleading~\cite{Wang-etal11}. For
example, in a sparse network,  the trivial classifier that labels
all samples as missing links can have a $99.99\%$ accuracy. In negative
link prediction, we aim to achieve high precision and recall over
negative links, defined in terms of the confusion matrix of a
classifier as shown in Table~\ref{tab:confusion}: $precision =
\frac{tp}{tp+fp}$ and $recall=\frac{tp}{tp+fn}$. Usually precision
and recall are combined into their harmonic mean, the F-measure.
Therefore we will adopt F1-measure as one metric for the performance
evaluation. As suggested in~\cite{Wang-etal11}, in some scenarios,
we put more emphasis on precision because the most challenging task
is to seek some negative links with high probability, even at the
price of increasing false negatives. Hence, we also report the
precision performance.

\begin{table}
    \vspace*{-0.15in}
\centering
\caption{Confusion Matrix of a Binary Classifier.}
\label{tab:confusion}
\begin{tabular}{|c|c|c|}
\hline
                              &True class = -1               &True class = 1\\ \hline
   Predicted class = -1       &true pos. (tp)         &false pos. (fp)        \\ \hline
   Predicted class = 1        &false neg. (fn)           &true neg. (tn)     \\
\hline
\end{tabular}
    \vspace*{-0.15in}
\end{table}

\subsection{Performance of Negative Link prediction}

In this subsection, we assess the proposed framework in terms of (a)
 the performance of NeLP with respect to baseline methods; and (b)
  the generalization of the proposed framework across social media
sites. For the first evaluation, we define the following baseline
methods:
\begin{itemize}
\item {\it Random}: This predictor randomly guesses the labels of samples.
As suggested in positive link prediction~\cite{Libe-etal07}, a random predictor should be used as a baseline method to
meaningfully demonstrate the performance significance of other predictors;
\item {\it sPath}: Observations in data analysis suggest that our ``enemies'' are
always close to us in the positive
network and {\it sPath} assigns negative links to pairs whose shortest path lengths is $L$;
\item {\it negIn}: Given the strong correlation between negative interactions and negative links,
{\it negIn} suggests negative links to these pairs with negative interactions;
\item {\it negInS}: after obtaining negative link candidates via {\it negIn}, {\it negInS} further refines these
candidates by performing a removing step and an adding step as shown in Algorithm~\ref{alg:negativesample}; and
\item {\it NeLP-negIn}: NeLP-negIn is a variant of the proposed NeLP framework.
Instead of using negative links suggested by {\it negInS} as NeLP, NeLP-negIn uses negative links found by {\it negIn}.
\end{itemize}

For parameterized methods, we report the best performance of each
baseline method. For NeLP, we set its parameters as
$\{C_p=1,C_n=0.5,C_b=0.1\}$ and $\{C_p=1,C_n=0.7,C_b=0.01\}$ in
Epinions and Slashdot, respectively. We empirically find that $f(x)
= 1 - \frac{1}{\log(1+x)}$ works well for the proposed framework.
More details about parameter sensitivity of NeLP will be discussed
in later subsections. The comparison results are demonstrated in
Table~\ref{tab:comparison}.

\begin{table}
    \vspace*{-0.15in}
\centering
\caption{Performance Comparison of Negative Link Prediction in Epinions and Slashdot.}
\label{tab:comparison}
\begin{tabular}{|c|c|c|c|c|}
\hline
\multirow{2}{*}{Algorithms}  & \multicolumn{2}{|c|}{Epinions} & \multicolumn{2}{|c|}{Slashdot} \\ \cline{2-5}
                              & F1     &Precision                              &F1     & Precision       \\\hline
{\it random}    &0.0005 &0.0002 &0.0008 &0.0004  \\\hline
{\it sPath}   &0.0040   &0.0075 &0.0090 &0.0172   \\\hline
{\it negIn}   &0.2826   &0.2097 &0.1986 &0.1483    \\\hline
{\it negInS}    &0.2893 &0.2124 &0.2072 &0.1524    \\\hline
{\it NeLP-negIn}    &0.3206 &0.2812 &0.2394 &0.2083  \\\hline
{\it NeLP}  &0.3242 &0.2861 &0.2441 &0.2139    \\\hline
\end{tabular}
    \vspace*{-0.15in}
\end{table}

We make the following observations:
\begin{itemize}
\item {\it sPath} obtains much better performance than random guessing, which further supports the hypothesis
that our ``enemies'' are close to us in the positive network;
\item {\it negIn} improves the performance significantly in both datasets. These results suggest the existence of
 correlation between negative interactions
and negative links;
\item by removing candidates suggested by {\it negIn} that do not satisfy status theory and adding candidates to make open triads
closure to satisfy status theory, {\it negInS} outperforms {\it
negIn}. For example, {\it negInS} gains $2.37\%$ and $1.45\%$
relative improvement in terms of F1-measure in Epinions and
Slashdot, respectively. These results indicate that status theory
can help us remove some noisy samples and add some useful samples
for training. These observations  can also be used to explain the
reason why the performance of NeLP based on negative links suggested
by {\it negInS} is better than that based on {\it negIn}; and
\item  the proposed framework always obtains the best performance. There are three important components of NeLP.
First, NeLP introduces $C_n$ to control errors in negative samples. Second, NeLP introduces $c_j$ to control the error in the sample
$x_j$, which is related to the number of negative interactions based
on our observations from data analysis. Third, NeLP introduces balance theory regularization to model balance theory, which
also allows us to include more samples in the classifier learning
process. More details about the effects of these components will be
discussed in a later subsection.
\end{itemize}

\begin{figure*}
    \begin{center}
      \subfigure[Epinions]{\label{fig:epinionscross}\includegraphics[scale=0.35]{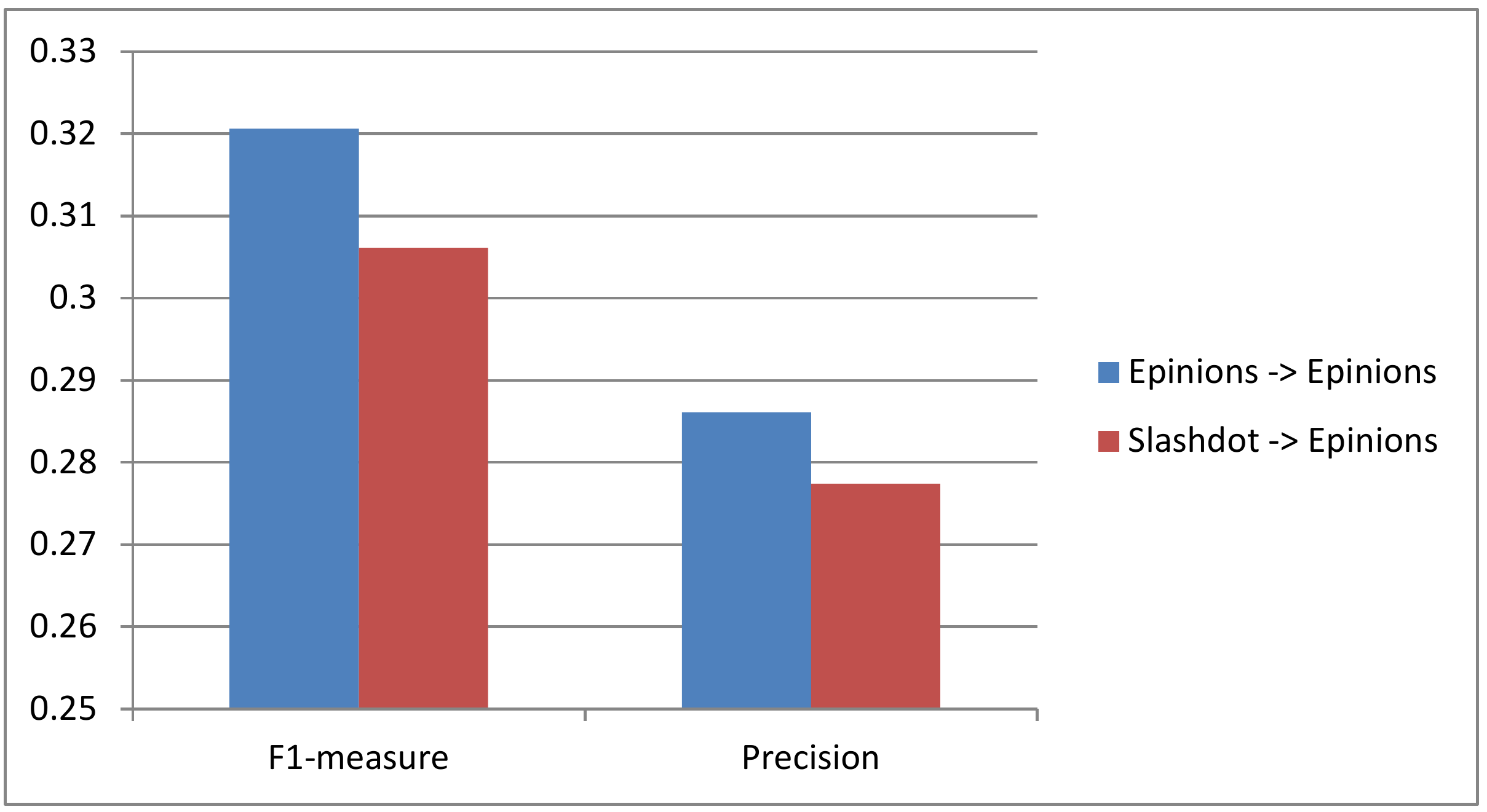}}
    \subfigure[Slashdot]{\label{fig:slashdotcross}\includegraphics[scale=0.35]{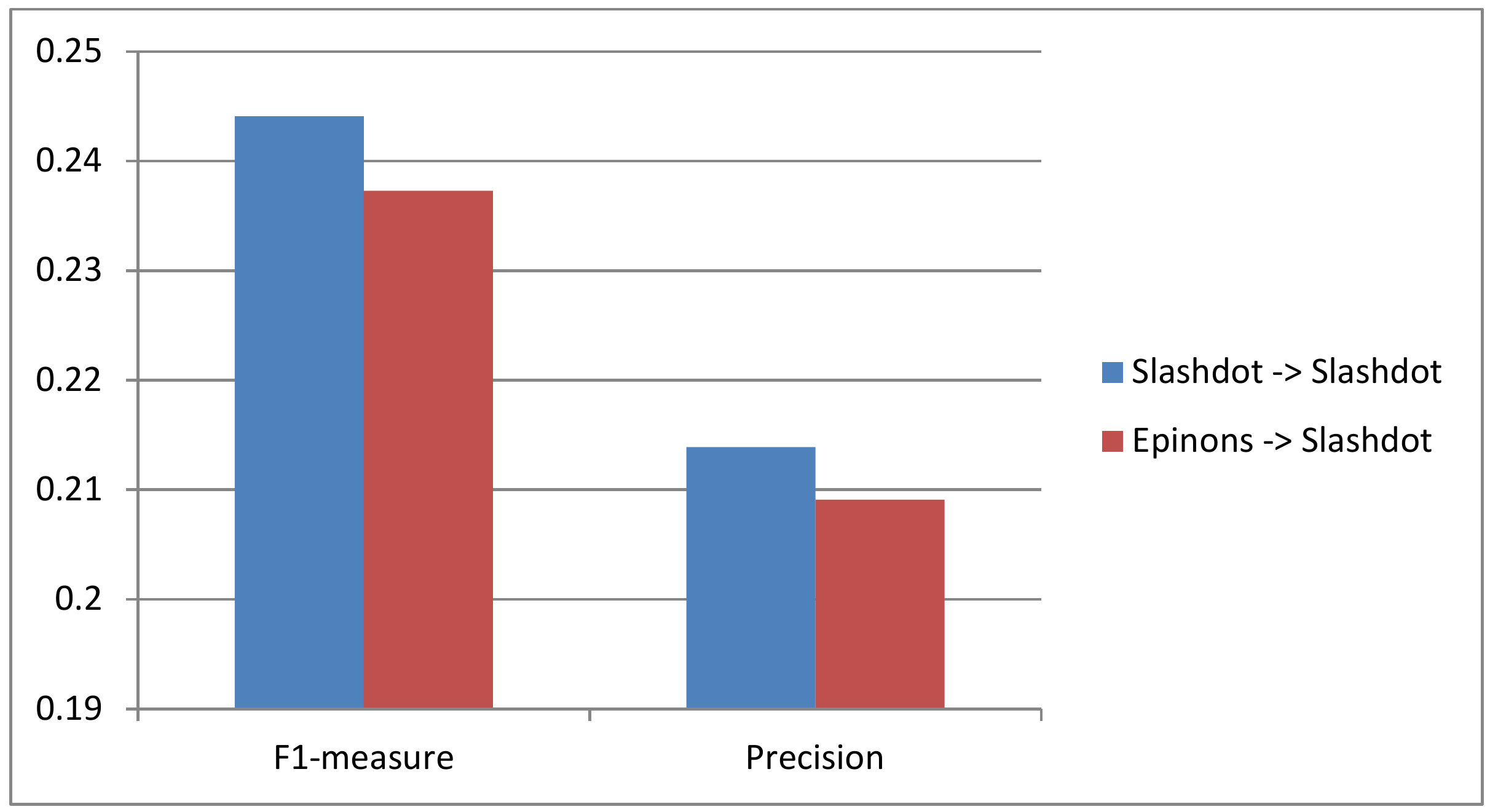}}
    \end{center}
  \vspace*{-0.15in}  
\caption{The Negative Link Prediction Performance across Epinions and Slashdot where $x \rightarrow y$ denotes training on $x$ and evaluating on $y$.}
\vspace*{-0.15in}
\label{fig:generalize}
\end{figure*}

Because the classifier learned by the proposed framework is based
on the same set of features extracted from pervasively available sources for most
social media sites, it is possible to generalize the classifier
learned in one site to other sites and we further investigate how
well the learned classifier generalizes
across social media sites. In particular, we evaluate the
performance of the classifier on Epinions (or Slashdot), which is
learned from Slashdot (or Epinions). The results are shown in
Figure~\ref{fig:generalize}. Note that in the figure $x \rightarrow
y$ denotes training on $x$ and evaluating on $y$. These results show
that there is very good generalization of  the classifier learned by
NeLP although there is remarkably little decrease in performance
regardless of which dataset is used for training. In summary,
compared to baseline methods, the proposed framework obtains
significant performance improvement, and it also has very good
generalization across social media sites.

\subsection{Component Analysis of NeLP}

In this subsection, we investigate the effects of various NeLP
components on its performance. In NeLP, we introduce $C_n$, $c_j$
and $C_b$ to control three components of NeLP. In particular,
 $C_n$ controls errors from negative samples, $c_j$ controls the
error from the negative sample $x_j$ and $C_b$ controls the
contribution from the balance theory regularization. By setting $C_p
= 1$ and varying different values of $C_n$, $c_j$ and $C_b$, we can
examine the impact of these components on the performance of NeLP.
The results of component analysis are shown in
Tables~\ref{tab:componentsepinions} and \ref{tab:componentsslashdot}
for Epinions and Slashdot, respectively.

\begin{table}
\centering
\caption{Component analysis for NeLP in Epinions.}
\label{tab:componentsepinions}
\begin{tabular}{|c|c|c|c|c|}
\hline
$C_n$ & $c_j$ & $C_b$ & F1-measure & Precision  \\ \hline
  0.5 &  $f(x) = 1 - \frac{1}{\log(1+x)}$ &0.1   &0.3242    &0.2861 \\ \hline
  1 &  $f(x) = 1 - \frac{1}{\log(1+x)}$   &0.1   &0.3188      &0.2793  \\ \hline
  0.5  &  f(x) = 1                        &0.1   &0.3067     &0.2612  \\ \hline
  0.5 &  $f(x) = 1 - \frac{1}{\log(1+x)}$ &0     &0.3084      &0.2686 \\ \hline
  1   &  f(x) = 1                         &0     &0.2992      &0.2342 \\ \hline
\end{tabular}
    \vspace*{-0.15in}
\end{table}

\begin{table}
\centering
\caption{Component Analysis of NeLP in Slashdot.}
\label{tab:componentsslashdot}
\begin{tabular}{|c|c|c|c|c|}
\hline
$C_n$ & $c_j$ & $C_b$ & F1-measure & Precision  \\ \hline
  0.7 &  $f(x) = 1 - \frac{1}{\log(1+x)}$ &0.01   &0.2441     &0.2139 \\ \hline
  1 &  $f(x) = 1 - \frac{1}{\log(1+x)}$   &0.01   &0.2403   &0.2094  \\ \hline
  0.7  &  f(x) = 1                   &0.01   &0.2287    &0.1902  \\ \hline
  0.7 &  $f(x) = 1 - \frac{1}{\log(1+x)}$ &0     &0.2347      &0.1972 \\ \hline
  1   &  f(x) = 1                      &0     &0.2213      &0.1722\\ \hline
\end{tabular}
    \vspace*{-0.15in}
\end{table}

The first row in each table represents the performance of NeLP with
all three components. We make the following observations about
different variations of NeLP in other rows of the table:
\begin{itemize}
\item in the second row, we set $C_n = 1$, which gives
equal weights to positive and negative samples. This approach
effectively eliminates the {\em differential} importance given to
errors from negative samples. The performance degrades, which
suggests that the errors of negative and positive samples should be
treated differently;
\item in the third row, we set $c_j=1$ instead of the reliability weight related to the number of negative interactions to eliminate the component controlling the error in the negative sample $x_j$. The performance reduces a lot.
For example, the precision reduces by $8.70\%$ and $11.08\%$ in
Epinions and Slashdot, respectively. These results support the
importance of the number of negative interactions to indicate the
reliability of negative samples;
\item in the fourth row, we set $C_b = 0$ to eliminate the contribution from the balance theory regularization.
and the performance is consistently worse than that with the balance
theory regularization. This illustrates the importance of the
balance theory regularization in the proposed NeLP framework; and
\item in the fifth row, we eliminate all these three components and the performance further degrades.
These results suggest that the three components contain
complementary information.
\end{itemize}

\subsection{Impact of  Balance Theory Regularization}

\begin{figure*}
    \begin{center}
      \subfigure[Epinions]{\label{fig:cbepinions}\includegraphics[scale=0.35]{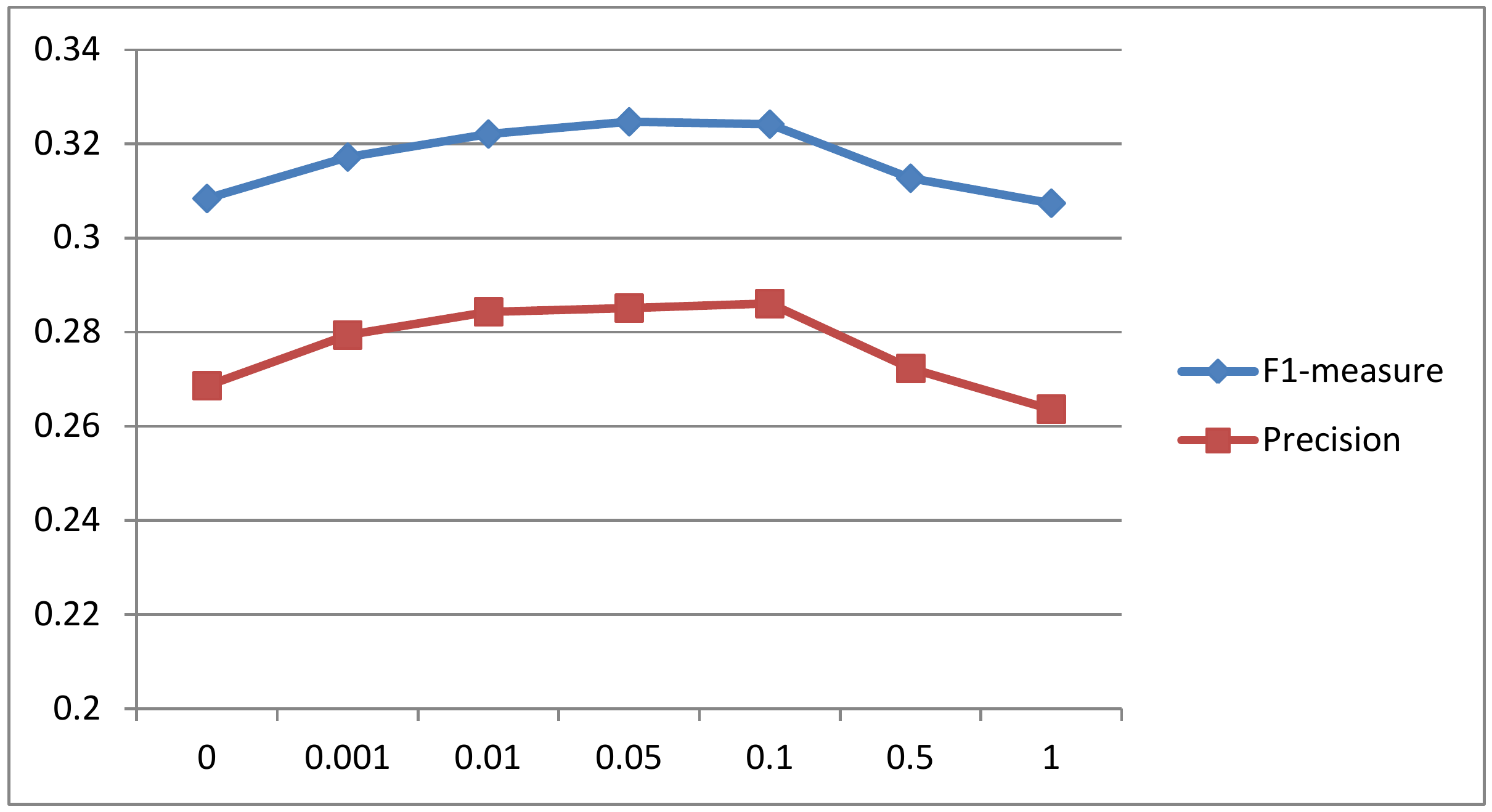}}
    \subfigure[Slashdot]{\label{fig:cbslashdot}\includegraphics[scale=0.35]{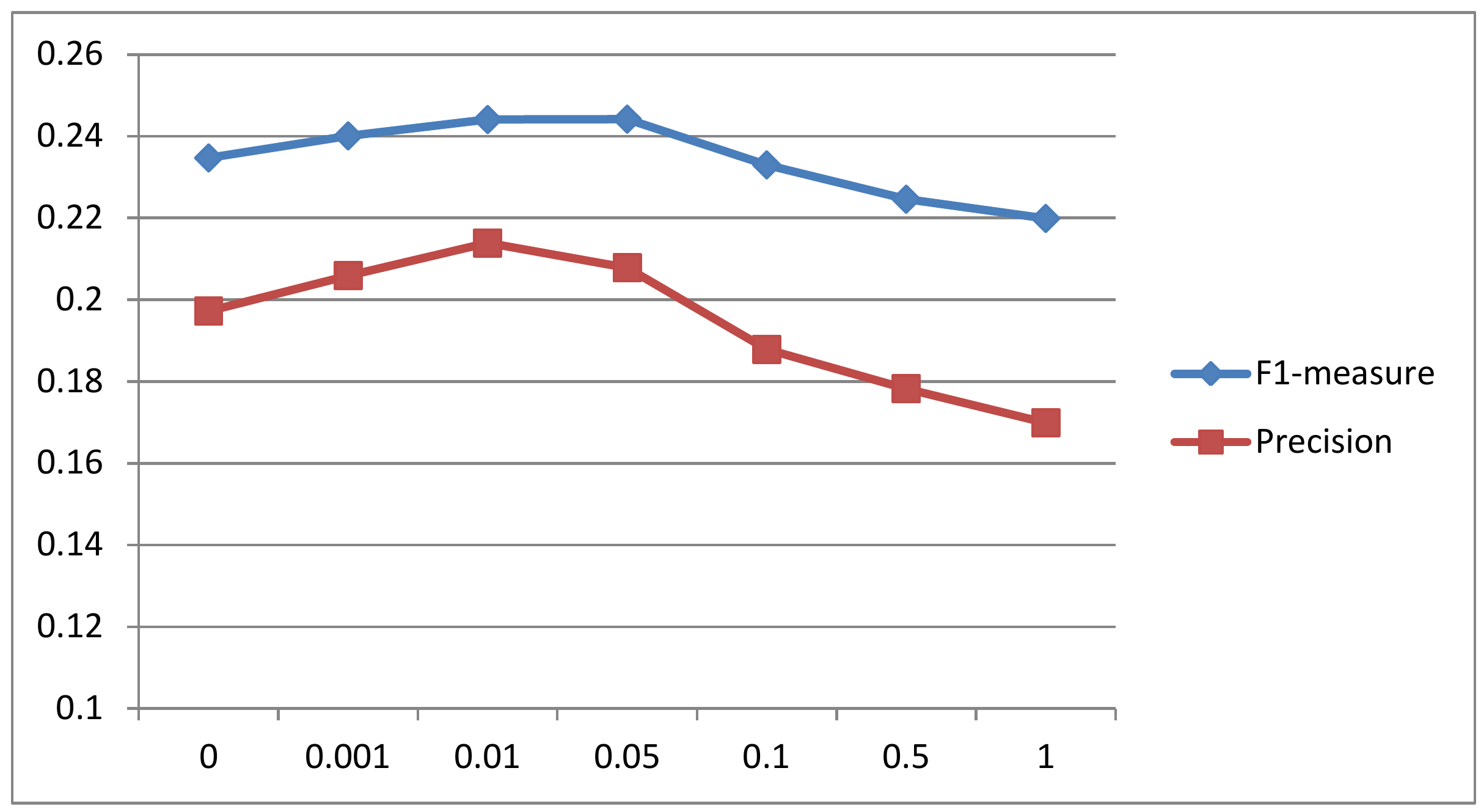}}
    \end{center}
 \vspace*{-0.15in}
\caption{The Impact of the Balance Theory Regularization on NeLP}
 \vspace*{-0.15in}
\label{fig:balancetheory}
\end{figure*}

The analysis of the previous subsection shows the importance of
balance theory regularization. In this subsection, we perform a more
detailed analysis  of  the impact of the balance theory
regularization on NeLP by showing how the performance varies with
the changes in the value of $C_b$. This parameter  controls the
contribution from the balance theory regularization. We vary the
values of $C_b$ as $\{0,0.001,0.01,0.05,0.1,0.5,1\}$ and the results
are shown in Figures~\ref{fig:cbepinions} and \ref{fig:cbslashdot}
for Epinions and Slashdot, respectively. In general, with increase
in $C_b$, the performance  first improves, peaks, and then drops
dramatically.

In particular, with the increase of $C_b$, we make the following observations:
\begin{itemize}
\item By increasing the value of $C_b$ from 0 to
 $0.001$, the performance increases significantly. For example, NeLP gains $4.02\%$ and $4.41\%$
  in terms of precision in Epinions and Slashdot, respectively. These results further support the
  importance of the balance theory regularization in the NeLP framework;
\item Within certain parameter ranges,  such as from 0.001 to 0.1 in Epinions, the performance is relatively stable.
 This property is practically useful because it makes it easier to set $C_b$; and
\item After certain values such as 0.05 in Slashdot, the performance decreases dramatically. A
large  value $C_b$ results in balance theory regularization
dominating the learning process at the expense of other factors.
\end{itemize}

In summary, the  parameter analysis on $C_b$ suggests that balance
theory regularization is important for the NeLP framework.

\section{Related Work}

In this section, we briefly review work which is related to
different variants of the link prediction problem.

\subsection{Positive Link Prediction}
Positive link prediction infers new positive links in the near
future based on a snapshot of a positive network. Existing methods
can be roughly divided into unsupervised methods and supervised
methods. Unsupervised methods are usually based on the topological
structure of the given positive network. In~\cite{Libe-Klei03},
several unsupervised link prediction algorithms are proposed, such
as Katz, Jaccard's coefficient and Adamic/Adar.
In~\cite{menon2011link}, several unsupervised algorithms based on
low-rank matrix factorization are proposed. There are usually two
steps for supervised methods. First, they extract features from
available sources to represent each pair of users and consider the
existence of positive links as labels. Second, they train a binary
classifier based on the representation with extracted features and
labels. In~\cite{lichtenwalter2010new}, the authors show several
advantages of supervised link prediction algorithms such as superior
performance, adaptation to different domains and variance reduction.
In~\cite{Wang-etal11}, the features extracted from human mobility
have very strong predictive power and can significantly improve the
positive link prediction performance.

\subsection{Positive and Negative Link Prediction}

Positive and negative link prediction infers new positive and
negative links by giving a snapshot of a signed network, which has
attracted increasing attention in recent years~\cite{borzymek2010trust}.
In~\cite{Guha-etal04}, an algorithm based on trust and distrust
propagation is proposed to predict trust and distrust relations.
In~\cite{Lesk-etal10}, local-topology-based features based on
balance theory are extracted to improve the performance of a
logistic regression classifier in signed relation prediction.
Features derived from longer cycles in signed networks can be used
to improve the positive and negative link prediction
performance~\cite{chiang2011exploiting}. In~\cite{hsieh2012low}, a
low-rank matrix factorization approach with generalized loss
functions is proposed to predict trust and distrust relations.

\subsection{Sign Prediction}

Sign prediction infers the sign of a given link.
In~\cite{yang2012friend}, user behavior of decision making can be
used to predict signs of a given unsigned network accurately. The
authors also show the importance of modeling balance theory and
status theory in the sign prediction problem. Tang et al. proposed a
framework to incorporate social theories such as status theory into
a machine learning model and infer the signs of links in a target
network by borrowing knowledge from a different source
network~\cite{tang2012inferring}. In~\cite{ye2013predicting}, the
authors use the transfer learning approach to leverage sign
information from an existing and mature signed network to predict
signs for a newly formed signed social network.

\section{Conclusion}

Research in signed network analysis suggests that negative links
have added value over positive links and they can potentially help
various social media services such as recommender systems. However,
most social media sites do not enable their users to specify
negative links. This makes the problem of negative link prediction
more challenging. In this paper, we investigate the problem of
negative link prediction. To preserve the generality of our
approach, we use  positive links and content-centric interactions as
sources to predict negative links because  these two sources are
pervasively available in social media. We first analyze the impact
of various social theories, such as balance theory and status
theory, on negative links.  Then we leverage these insights to
provide a principled way to exploit positive links and
content-centric interactions. Finally, we propose the  NeLP
framework for negative link prediction. Experimental results on two
real-world social media datasets demonstrate the effectiveness and
generalization of the proposed framework. Further experiments
illustrate the impact of various model components.

There are several interesting directions needing further
investigation. First, negative links predicted by the proposed
framework may benefit various social media applications such as
positive link prediction and recommender systems. Therefore,  we
plan to investigate how to incorporate the proposed framework into
these applications to improve their performance. Second, in addition
to positive links and content-centric interactions, the
user-generated content is also pervasively available in social
media. We would like to investigate whether user generated content
is useful and how to exploit it for the negative link prediction
problem. Finally the constructed labels in the studied problem may
be noisy; hence we choose a noise-tolerant support vector machine as
the basic algorithm for the proposed framework. Learning with noisy
labels has been extensively studied in the machine learning
community~\cite{frenay2013classification} and we will experiment
with the use of other noise-tolerant algorithms as basic algorithms
for the problem of negative link prediction.

\section*{Acknowledgments}

This material is based upon work supported by, or in part by, the U.S. Army Research Office (ARO) under contract/grant
number 025071, the Office of Naval Research(ONR) under grant number N000141010091, and the Army Research Laboratory and was accomplished under Cooperative Agreement Number W911NF-09-2-0053. The views and conclusions contained in this document are those of the authors and should not be interpreted as representing the official policies, either expressed or implied, of the Army Research Laboratory or the U.S. Government. The U.S. Government is authorized to reproduce and distribute reprints for Government purposes notwithstanding any copyright notation here on.

\bibliographystyle{abbrv}

\begin{thebibliography}{10}

\bibitem{belkin2005manifold}
M.~Belkin, P.~Niyogi, and V.~Sindhwani.
\newblock On manifold regularization.
\newblock In {\em Proceedings of the Tenth International Workshop on Artificial
  Intelligence and Statistics (AISTAT 2005)}, pages 17--24, 2005.

\bibitem{Cart-etal56}
D.~Cartwright and F.~Harary.
\newblock Structural balance: a generalization of heider's theory.
\newblock {\em Psychological Review}, 63(5):277, 1956.

\bibitem{chiang2013prediction}
K.-Y. Chiang, C.-J. Hsieh, N.~Natarajan, A.~Tewari, and I.~S. Dhillon.
\newblock Prediction and clustering in signed networks: A local to global
  perspective.
\newblock {\em arXiv preprint arXiv:1302.5145}, 2013.

\bibitem{chiang2011exploiting}
K.-Y. Chiang, N.~Natarajan, A.~Tewari, and I.~S. Dhillon.
\newblock Exploiting longer cycles for link prediction in signed networks.
\newblock In {\em Proceedings of the 20th ACM international conference on
  Information and knowledge management}, pages 1157--1162. ACM, 2011.

\bibitem{cristianini2000introduction}
N.~Cristianini and J.~Shawe-Taylor.
\newblock {\em An introduction to support vector machines and other
  kernel-based learning methods}.
\newblock Cambridge university press, 2000.

\bibitem{frenay2013classification}
B.~Fr{\'e}nay, M.~Verleysen, et~al.
\newblock Classification in the presence of label noise: a survey.
\newblock {\em IEEE Transactions on Neural Networks and Learning Systems},
  pages 25--17, 2013.

\bibitem{Guha-etal04}
R.~Guha, R.~Kumar, P.~Raghavan, and A.~Tomkins.
\newblock Propagation of trust and distrust.
\newblock In {\em Proceedings of the 13th international conference on World
  Wide Web}, pages 403--412. ACM, 2004.

\bibitem{hardin2004distrust}
R.~Hardin.
\newblock Distrust: Manifestations and management.
\newblock {\em Distrust}, 8:3--33, 2004.

\bibitem{heider1946attitudes}
F.~Heider.
\newblock Attitudes and cognitive organization.
\newblock {\em The Journal of psychology}, 21(1):107--112, 1946.

\bibitem{hsieh2012low}
C.-J. Hsieh, K.-Y. Chiang, and I.~S. Dhillon.
\newblock Low rank modeling of signed networks.
\newblock In {\em Proceedings of the 18th ACM SIGKDD international conference
  on Knowledge discovery and data mining}, pages 507--515. ACM, 2012.

\bibitem{kunegis2009slashdot}
J.~Kunegis, A.~Lommatzsch, and C.~Bauckhage.
\newblock The slashdot zoo: mining a social network with negative edges.
\newblock In {\em Proceedings of the 18th international conference on World
  wide web}, pages 741--750. ACM, 2009.

\bibitem{kunegis2013added}
J.~Kunegis, J.~Preusse, and F.~Schwagereit.
\newblock What is the added value of negative links in online social networks?
\newblock In {\em Proceedings of the 22nd international conference on World
  Wide Web}, pages 727--736. International World Wide Web Conferences Steering
  Committee, 2013.

\bibitem{Lesk-etal10}
J.~Leskovec, D.~Huttenlocher, and J.~Kleinberg.
\newblock Predicting positive and negative links in online social networks.
\newblock In {\em Proceedings of the 19th international conference on World
  wide web}, 2010.

\bibitem{Lesk-etalchi10}
J.~Leskovec, D.~Huttenlocher, and J.~Kleinberg.
\newblock Signed networks in social media.
\newblock In {\em Proceedings of the 28th international conference on Human
  factors in computing systems}, pages 1361--1370. ACM, 2010.

\bibitem{Libe-Klei03}
D.~Liben-Nowell and J.~Kleinberg.
\newblock The link prediction problem for social networks.
\newblock In {\em Proceedings of 12th International Conference on Information
  and Knowledge Management}, 2003.

\bibitem{Libe-etal07}
D.~Liben-Nowell and J.~Kleinberg.
\newblock The link-prediction problem for social networks.
\newblock {\em Journal of the American society for information science and
  technology}, 58(7):1019--1031, 2007.

\bibitem{lichtenwalter2010new}
R.~N. Lichtenwalter, J.~T. Lussier, and N.~V. Chawla.
\newblock New perspectives and methods in link prediction.
\newblock In {\em Proceedings of the 16th ACM SIGKDD international conference
  on Knowledge discovery and data mining}, pages 243--252. ACM, 2010.

\bibitem{ma2009learning}
H.~Ma, M.~R. Lyu, and I.~King.
\newblock Learning to recommend with trust and distrust relationships.
\newblock In {\em Proceedings of the third ACM conference on Recommender
  systems}, pages 189--196. ACM, 2009.

\bibitem{menon2011link}
A.~K. Menon and C.~Elkan.
\newblock Link prediction via matrix factorization.
\newblock In {\em Machine Learning and Knowledge Discovery in Databases}, pages
  437--452. Springer, 2011.

\bibitem{papadopoulos2012community}
S.~Papadopoulos, Y.~Kompatsiaris, A.~Vakali, and P.~Spyridonos.
\newblock Community detection in social media.
\newblock {\em Data Mining and Knowledge Discovery}, 24(3):515--554, 2012.

\bibitem{tang2013social}
J.~Tang, X.~Hu, and H.~Liu.
\newblock Social recommendation: a review.
\newblock {\em Social Network Analysis and Mining}, 3(4):1113--1133, 2013.

\bibitem{tang2012feature}
J.~Tang and H.~Liu.
\newblock Feature selection with linked data in social media.
\newblock In {\em SDM}, pages 118--128. SIAM, 2012.

\bibitem{tang2012inferring}
J.~Tang, T.~Lou, and J.~Kleinberg.
\newblock Inferring social ties across heterogenous networks.
\newblock In {\em Proceedings of the fifth ACM international conference on Web
  search and data mining}, pages 743--752. ACM, 2012.

\bibitem{victor2009trust}
P.~Victor, C.~Cornelis, M.~De~Cock, and A.~Teredesai.
\newblock Trust-and distrust-based recommendations for controversial reviews.
\newblock In {\em Web Science Conference (WebSci'09: Society On-Line)}, number
  161, 2009.

\bibitem{Wang-etal11}
D.~Wang, D.~Pedreschi, C.~Song, F.~Giannotti, and A.~Barab{\'a}si.
\newblock Human mobility, social ties, and link prediction.
\newblock In {\em KDD}, pages 1100--1108. ACM, 2011.

\bibitem{yang2012friend}
S.-H. Yang, A.~J. Smola, B.~Long, H.~Zha, and Y.~Chang.
\newblock Friend or frenemy?: predicting signed ties in social networks.
\newblock In {\em SIGIR}, pages 555--564. ACM,
  2012.

\bibitem{ye2013predicting}
J.~Ye, H.~Cheng, Z.~Zhu, and M.~Chen.
\newblock Predicting positive and negative links in signed social networks by
  transfer learning.
\newblock In {\em WWW}, pages 1477--1488, 2013.

\bibitem{zhu2003semi}
X.~Zhu, Z.~Ghahramani, J.~Lafferty, et~al.
\newblock Semi-supervised learning using gaussian fields and harmonic
  functions.
\newblock In {\em ICML}, volume~3, pages 912--919, 2003.

\end{thebibliography}

\end{document}